\documentclass[journal=ancac3,manuscript=article]{achemso}
\pdfoutput=1 

\usepackage{graphicx}
\usepackage{subcaption}
\usepackage{amsmath,bm}
\usepackage{siunitx,physics}
\graphicspath{{images/}}

\usepackage[T1]{fontenc}

\author{Fred Fu}
\author{Nasser Mohieddin Abukhdeir}
\affiliation{Department of Chemical Engineering, University of Waterloo, Waterloo, Ontario N2L 3G1, Canada}
\alsoaffiliation{Department of Physics and Astronomy, University of Waterloo, Waterloo, Ontario N2L 3G1, Canada}
\alsoaffiliation{Waterloo Institute for Nanotechnology, University of Waterloo, Waterloo, Ontario N2L 3G1, Canada}
\email{nmabukhdeir@uwaterloo.ca}

\title[Dynamics of Nematic Spheroids]{Formation and field-driven dynamics of nematic spheroids}


\begin{document}

\begin{tocentry}
	\centering
	\includegraphics[height=3.5cm,keepaspectratio]{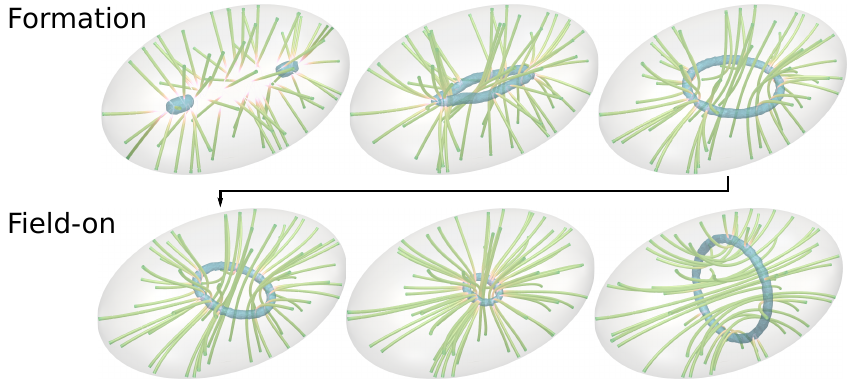}
\end{tocentry}

\begin{abstract}
    Emerging technologies based on liquid crystal (LC) materials increasingly leverage the presence of nanoscale defects, unlike the canonical application of LCs -- LC displays.
    The inherent nanoscale characteristics of LC defects present both significant opportunities and barriers for the application of this fascinating class of materials.
    Simulation-based approaches to the study of the effects of confinement and interface anchoring conditions on LC domains has resulted in significant progress over the past decade, where simulations are now able to access experimentally-relevant micron-scales while simultaneously capturing nanoscale defect structures.
    In this work, continuum simulations were performed in order to study the dynamics of micron-scale nematic LC droplets of varying spheroidal geometry.
    Nematic droplets are one of the simplest inherently defect-containing LC structures and are also relevant to polymer-dispersed LC-based ``smart'' window technology.
    Simulation results include nematic phase formation and external field-switching dynamics of droplets ranging in shape from oblate to prolate.
    Results include both qualitative and quantitative insight into the complex coupling of nanoscale defect dynamics and structure transitions to micron-scale  reorientation.
    Dynamic mechanisms are presented and related to structural transitions in LC defects present in the droplet.
    Droplet-scale metrics including order parameters and response times are determined for a range of experimentally-accessible electric field strengths.
    These results have both fundamental and technological relevance, in that increased understanding of LC dynamics in the presence of defects is a key barrier to continued advancement in the field.
\end{abstract}

Liquid crystals (LCs) are materials which exhibit properties characteristic of both disordered liquids and crystalline solids.
Their anisotropic nature imparts unique optical properties and makes them susceptible to external fields.
These properties have resulted in a wide array of electro-optical applications, such as liquid crystal displays (LCDs).
However, unlike LCDs, which are designed using uniform defect-free domains, next-generation LC-based technologies are increasingly leveraging the presence of nanoscale topological defects.
These emerging technologies include tunable photonics based on blue LC phases  \cite{Coles2010}, molecular self-assembly \cite{Wang2015}, and bistable optical devices \cite{Serra2011}.
Consequently, understanding and predicting defect-enabled LC phenomena is a key barrier to continued advancements, both fundamental and technological.
Theoretical and computational research is necessary to overcome this barrier due to the nanoscale lengths and times associated with LC structure and dynamics, which are currently inaccessible via experimental methods.

\begin{figure}
	\centering
    \begin{subfigure}[b]{0.8\linewidth}
        \centering
        \includegraphics[width=0.4\linewidth]{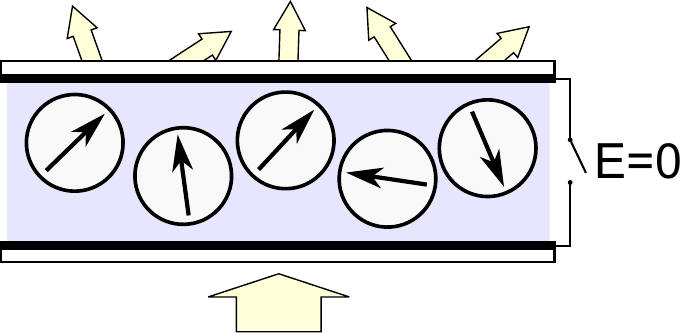}
        \quad
        \includegraphics[width=0.4\linewidth]{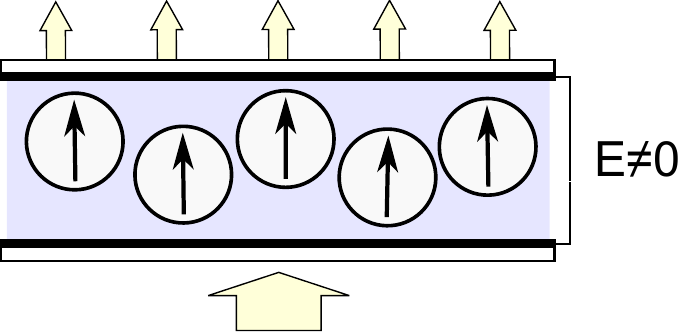}
        \caption{}\label{fig:pdlc-op}
    \end{subfigure}
    \begin{subfigure}[b]{0.5\linewidth}
        \centering
        \includegraphics[width=0.4\linewidth]{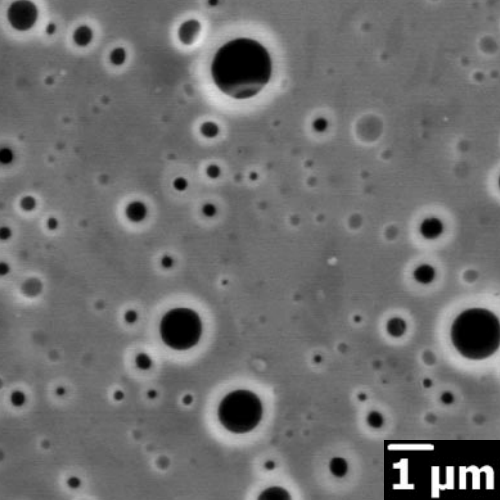}
        \qquad
        \includegraphics[width=0.4\linewidth]{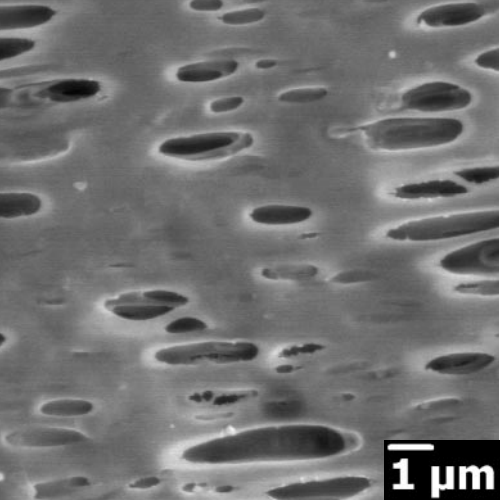}
        \caption{}\label{fig:stretch}
    \end{subfigure}
    \caption{(a) Schematic of the operation of a PDLC-based ``smart'' window where light is scattered by (left) randomly oriented nematic droplets in the absence of an electric field (translucent mode), which when exposed to an external field (right) are aligned in the direction normal to film (transparent mode). (b) SEM images of an (left) unstretched and (right) uniaxially stretched PDLC film where the resulting droplet shape is anisometric. Reproduced with permission from ref. \citenum{Amimori2003}.}
    \label{fig:pdlcs}
\end{figure}

One of the simplest inherently defect-containing structures is an LC droplet.
When LC material is confined in this way, a frustrated domain with significant spatial variation in LC order can emerge.
This so-called LC ``texture'' can differ depending on LC/solid anchoring conditions, domain shape, and LC material properties \cite{Serra2016}.
LC droplets play a major role in polymer-dispersed liquid crystal (PDLC) films, which are typically fabricated through a ``bottom-up'' process which results in nano-to-microscale LC domains dispersed in a polymer matrix.
PDLCs are optical functional materials which exhibit an optical response when subjected to thermal or external field actuation (Figure \ref{fig:pdlc-op}), introducing complex dynamics and constraints on response and relaxation times between equilibrium states.
PDLC films have traditionally been used for optical light shutter technology \cite{Drzaic1988}, in which LC domains are micron-sized.
More recently however, PDLCs incorporating nano-sized domains have been incorporated into novel applications such as holographic PDLC (H-PDLC) lasers and tunable microlens arrays \cite{Bunning2000,Coles2010}.

PDLC performance is governed by a variety of material and operating parameters, including LC defect-mediated structure and dynamics due to the topological constraints on LC order resulting from spheroidal confinement.
It has been more than two decades since Drzaic \cite{Drzaic1988} found that domain shape, specifically anisometry, strongly affects device performance \cite{Wu1989}.
Since then, it has been demonstrated that this anisometry can be directly controlled through various means \cite{Aphonin1993,Klosowicz2005}, the simplest of which is by uniaxial mechanical stretching of the PDLC film to produce highly prolate spheroidal domains (Figure~\ref{fig:stretch}) \cite{Amimori2003, Amimori2005}.

While a significant body of past mesoscale simulation work exists for cylindrical nematic capillaries \cite{DeLuca2007,Rey2007} and spherical droplets \cite{Chan1997,Li1999,Armas-Perez2015}, elliptic or ellipsoidal domains have been far less studied \cite{Bharadwaj2000,Chan1999,Rudyak2013,Khayyatzadeh2015}.
Furthermore, of this work, most use theoretical models which are unable to accurately capture nematic defects and phase transition.
Only recently have simulations been performed which capture nematic dynamics \cite{Abukhdeir2016a}, as opposed to just determining equilibrium states.
As a result, while past research has provided some insight into the nanoscale defect structure present in these domains, as of yet there have been no simulations of the dynamics of nematic spheroids on relevant length and timescales.
Thus our aim is to predict the dynamic mechanisms involved in the formation, field switching, and relaxation of nematic spheroids, such as those present in PDLC-based devices.
This objective has both fundamental and technological relevance in that these dynamic mechanisms are both poorly understood and necessary for advancement of this technology.
From a fundamental perspective, PDLCs provide ideal templates for the study of nanoscale defect behaviour in confined LC domains.
From a technological perspective, significant improvement in the performance of PDLCs as electro- and thermo-optical functional materials is required for their broader commercialization.

\section{NEMATIC PROPERTIES AND DYNAMIC MODEL}

LCs include a wide variety of phases, referred to as mesophases, with the simplest mesophase being the nematic phase.
Nematics exhibit not only translational disorder like a traditional liquid but also long-range orientational order, as shown by their tendency to self-align at the molecular scale.
Technological applications of nematic LCs, such as LCDs, mainly involve domains that are at or close to hydrostatic equilibrium which is likely due to the significant complexity of accounting for LC hydrodynamics \cite{Rey2002,Yang2010}.
Dynamics within this regime are referred to as reorientation dynamics, in which the orientation of individual LC molecules, or mesogens, evolve in response to thermodynamic or external stimuli.
This LC orientation can be described using the continuum Landau--de Gennes model of the nematic phase\cite{deGennes1995}, which introduces a symmetric traceless tensor order parameter called the alignment tensor\cite{Sonnet1995},
\begin{equation}
    Q_{ij} = S(n_i n_j - \frac{1}{3}\delta_{ij}) + P(m_i m_j - l_i l_j)
\end{equation}
which approximates the local orientational distribution function of the mesogens at each point in space.
The alignment tensor $\bm{Q}$ may be decomposed into its eigenvalues and eigenvectors, which describe the local orientational axis or nematic director $\bm{n}$, the uniaxial scalar order parameter $S$, and the biaxial scalar order parameter $P$ (and its associated axes, given by $\bm{m}$ and $\bm{l}$).
For a nematic domain, $S=P=0$ corresponds to the isotropic phase (a traditional disordered liquid), while $0 < S < 1$ and $P=0$ corresponds to the (uniaxial) nematic phase where a higher value of $S$ corresponds to greater alignment.
Biaxial orientational ordering occurs when both $S$ and $P$ are non-zero.

\begin{figure}
    \begin{subfigure}[b]{0.3\linewidth}
        \centering
        \includegraphics[width=\linewidth]{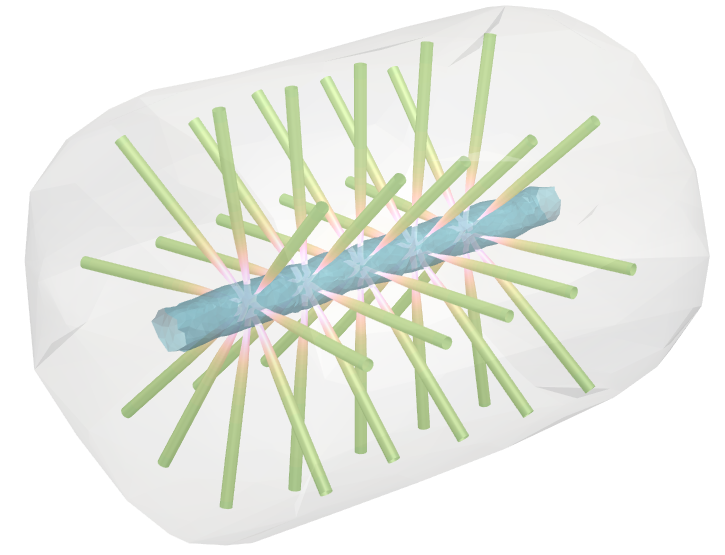}
        \caption{}\label{fig:defectline}
    \end{subfigure}
    \begin{subfigure}[b]{0.3\linewidth}
        \centering
        \includegraphics[width=0.9\linewidth]{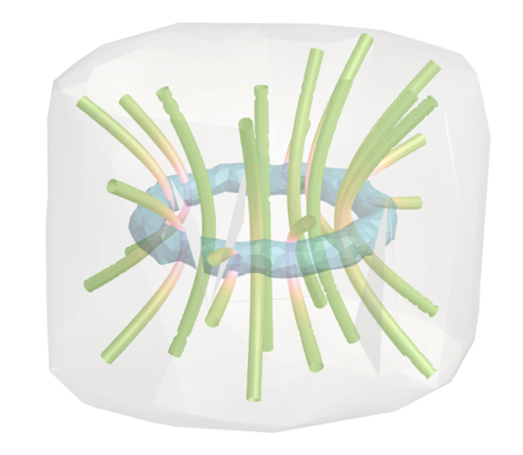}
        \caption{}\label{fig:defectloop}
    \end{subfigure}

    \caption{Schematics of (a) $+1$ and (b) $+\frac{1}{2}$ disclination lines using a combination of hyperstreamlines to indicate nematic orientation and an isosurface indicating the nanoscale defect ``core'' region.}
    \label{fig:defects}
\end{figure}

The majority of past simulation-based research on nematic LCs neglects variations in $S$, which results in a simplified model involving only the nematic director $\vb{n}$ \cite{Frank1958,Zumer1992},
\begin{multline}
    f(\vb{n},\grad{\vb{n}})=f_{0}+\frac{1}{2}k_{11}(\div{\vb{n}})^{2} + \frac{1}{2}k_{22}(\vb{n}\vdot\curl{\vb{n}} )^{2} +  \frac{1}{2}k_{33}(\vb{n}\cross\curl{\vb{n}})^{2} \\ - \frac{1}{2}k_{24}\div{(\vb{n}(\div{\vb{n}})+\vb{n}\cross\curl{\vb{n}})^{2})}
\end{multline}
which includes elastic energy terms that quantify the nematic response to orientational deformations of splay $k_{11}$, twist $k_{22}$, bend $k_{33}$, and saddle-splay $k_{24}$.
Many past simulation studies of elliptic nematic capillaries and ellipsoidal droplets\cite{Bharadwaj2000,Chan1997,Rudyak2013} use this simplified model despite its inability to accurately capture nanoscale defects in nematic order, called disclinations \cite{Sonnet1995}.
Disclinations correspond to singularities in the nematic director $\bm{n}$, and are therefore also regions of high biaxial nematic order ($S, P > 0$), as opposed to isotropic regions of disorder ($S=P=0$).
Figure~\ref{fig:defects} shows schematics of the two main types of disclinations relevant to nematic droplets: $+1$ line and $+\frac{1}{2}$ loop disclinations.

In contrast, the Landau--de Gennes model (see Methods section) is able to accurately capture both the presence of disclinations in nematic domains as well as their dynamics.
A review of recent studies using this model to simulate nematic dynamics may be found in ref. \citenum{Abukhdeir2016a}.
However, there are two major shortcomings of past simulation studies of nematic LC droplets.
The first is the widely-used single elastic constant approximation, where it is assumed that $k_{11} = k_{22} = k_{33}$ and $k_{24}=0$, despite the fact that these constants can widely differ, even for commonly studied LCs\cite{Luckhurst2001} and may significantly affect simulation outcomes\cite{Wincure2006}.
The second shortcoming is the sparsity of dynamic simulations, which can offer greater insight than simply solving for equilibrium nematic textures.
In this study, material parameters are used that correspond to the 4'-pentyl-4-cyanobiphenyl (5CB), a well-characterized nematic LC.
The domain is assumed to be isothermal, at hydrostatic equilibrium ($\vb{v} = \vb{0}$), and fluctuations in nematic order are neglected.
These assumptions are consistent with past simulations\cite{Ravnik2009,Tomar2012} except that simplifications of nematic elasticity are not made in this work.

Finally, in addition to nematic elasticity, interfacial surface anchoring effects arising from factors such as PDLC composition \cite{Kitzerow1994} must be considered.
Surface anchoring may result in a preferred nematic director $\vb{n}$ at the droplet interface and also the enhancement of nematic ordering $S > S_{0}$, where $S_0$ is the value of $S$ at thermodynamic equilibrium.
In this study, the case of homeotropic anchoring is investigated, in which $\vb{n} \parallel \vb{k}$ is energetically preferred, where $\vb{k}$ is the unit normal vector to the LC droplet surface.
Several experimental studies of PDLC dynamics have been performed under these conditions \cite{Mei2000,Xie2004,Prischepa2005,Boussoualem2014}.

In order to study the formation and field-driven dynamics of spheroidal nematic domains relevant to electro-optical applications of PDLCs, simulations were performed of nematic spheroids with fixed volume corresponding to an initial ``unstretched'' sphere of diameter $\SI{500}{\nano\meter}$.
To emulate stretching of the droplets, the initial sphere was consistently elongated or contracted along a single direction.
Droplet aspect ratio $R$ is defined as the length ratio between the axis of elongation/contraction and the remaining (equivalent) axes of the spheroid, resulting in oblate droplets for $R < 1$ and prolate droplets for $R > 1$.
Various aspect ratio $R$ domains were simulated in the interval $[0.5, 2]$ based upon experimental evidence regarding the expected variation in droplet shape \cite{Drzaic1988,Mei1998,Erdmann1989,Aphonin1993}.

These simulations were performed in three stages: (i) formation of the nematic phase from an initially disordered (high temperature) phase, (ii) application of an electric field corresponding to the ``on'' (transparent) state of a PDLC film, and (iii) relaxation resulting from release of the electric field, corresponding to the ``off'' (translucent) state of a PDLC film.
For the formation dynamics simulations, heterogeneous nucleation of the nematic phase at the solid/LC interface was assumed based on recent experimental observations \cite{Aya2011}.
For the field dynamics simulations, a range of experimentally accessible electric field strengths up to $\SI{14}{\volt\per\micro\meter}$ were applied.
Further details of the nematic dynamics model, numerical methods, and auxiliary conditions used in these simulations may be found in the Methods section.

\section{FORMATION FROM DISORDERED PHASE}\label{sec:res:formation}

Formation dynamics simulations were initially performed for oblate spheroids of aspect ratio $R \in [0.5, 1)$.
This geometry can be considered a rotational extrusion of a two-dimensional ellipse about its minor axis.
It is therefore comparable to previous simulations of nematic elliptic capillaries \cite{Khayyatzadeh2015}, in which a sequence of three different growth regimes were identified during droplet formation: (i) free growth, (ii) defect formation, and (iii) bulk relaxation.
The free growth regime consists of the stable nematic ``shell'' growing into an unstable isotropic phase, with the bulk nematic orientation being commensurate with the homeotropic surface anchoring conditions.
Next, the defect formation regime involves the impingement of the nematic-isotropic interface on itself.
This resultes in the simultaneous formation of a pair of $+\frac{1}{2}$ disclination lines along the major axis of the elliptic cross-section of the capillary.
Finally, during bulk relaxation, the domain as a whole relaxes to its equilibrium state through simultaneous disclination motion and bulk reorientation.

\begin{figure}
	\centering
	\includegraphics[width=0.98\linewidth]{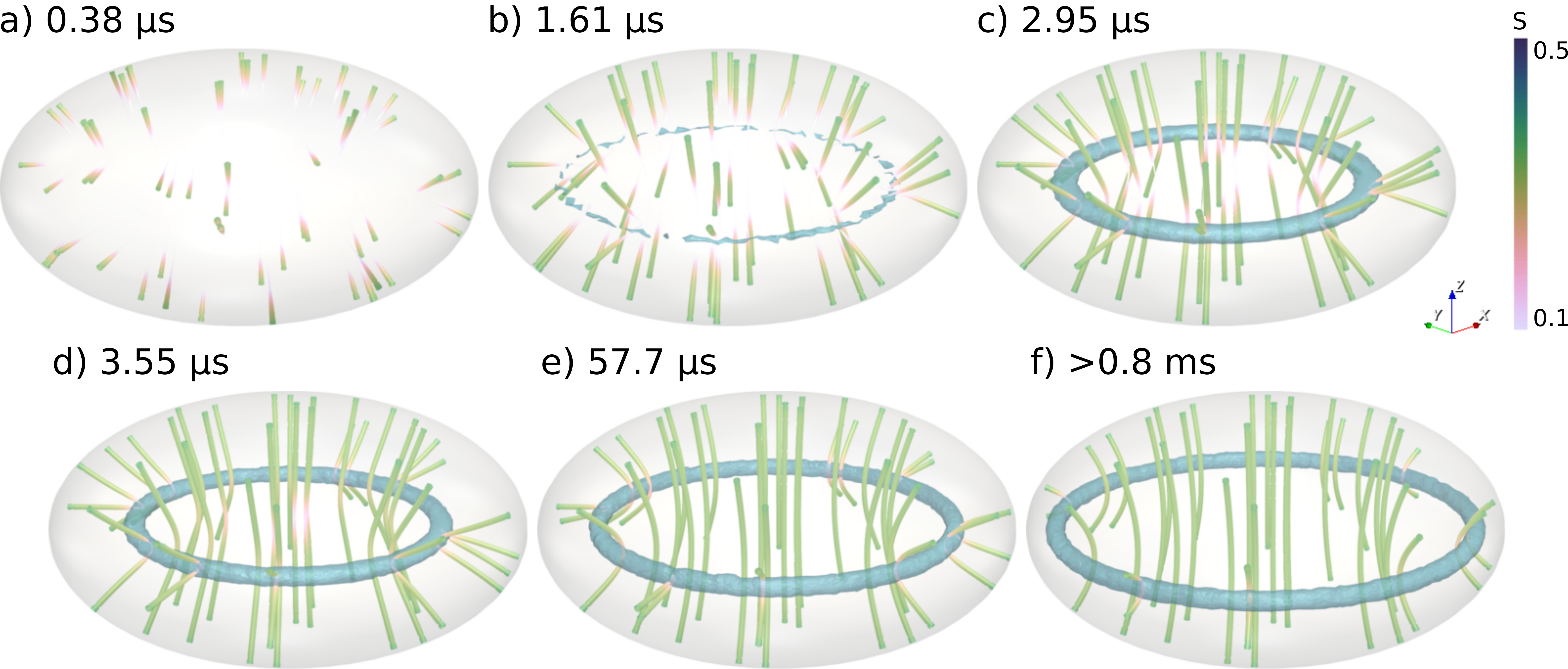}
	\caption{Simulation visualizations of the formation process of an oblate ($R = 0.5$) nematic droplet from an initially isotropic (disordered) state.  Hyperstreamlines colored by the magnitude of the uniaxial nematic scalar order parameter $S$ are used to visualize nematic orientation (alignment tensor) and isosurfaces indicate nanoscale defect ``core'' regions.}
	\label{fig:formation_oblate}
\end{figure}

The simulation results of the formation process for a $R=0.5$ oblate droplet are shown in Figure \ref{fig:formation_oblate}.
The same set of growth regimes can be identified, starting with the initial free growth of the stable nematic boundary layer into the central unstable isotropic region  (Figures~\ref{fig:formation_oblate}a--b).
As free growth proceeds, the curvature of the isotropic/nematic interface increases in the focal regions of the spheroid and simultaneously the interface velocity decreases.
This critical slowing down of the nematic/isotropic interface \cite{Khayyatzadeh2015} may be explained by an approximation of the interface velocity $v$ \cite{Wincure2006},
\begin{equation}
    \beta v = \mathcal{C} - \Delta F
\end{equation}
where $\beta$ is an effective viscosity term, $\Delta F$ is the difference in energy between the nematic and isotropic phases, and $\mathcal{C}$ the capillary force.
For an isothermal domain, $\Delta F$ is constant and, in the absence of curvature of the interface ($\mathcal{C} = 0$), the model predicts constant interface velocity $v$.
While the isotropic/nematic interfaces in the equatorial region of the droplet are able to grow inwards with minimal increase in interface curvature, this is not the case for focal regions of the interface.
As the radii of curvature of the interfaces in this region approach the nematic coherence length $\lambda_n\approx \SI{10}{nm}$ \cite{deGennes1995}, the capillary force $\mathcal{C}$ approaches the difference in free energy resulting from the transition $\Delta F$ and $v\rightarrow 0$.
At this point, the free growth regime transitions to the defect formation regime.

Figures~\ref{fig:formation_oblate}c--d show the defect formation regime dynamics.
Simultaneously, a $+\frac{1}{2}$ disclination loop forms in the focal region through a interface-driven defect ``shedding'' mechanism \cite{Wincure2007a} and the isotropic/nematic fronts in the equatorial region impinge.
This is followed by the bulk relaxation regime where the droplet texture relaxes through bulk reorientation and the disclination loop expands towards the focal boundaries.
As expected, the formation process of oblate nematic droplets is analogous to that of nematic elliptic capillaries due to their geometric similarities.

\begin{figure}
   	\centering
   	\includegraphics[width=0.98\linewidth]{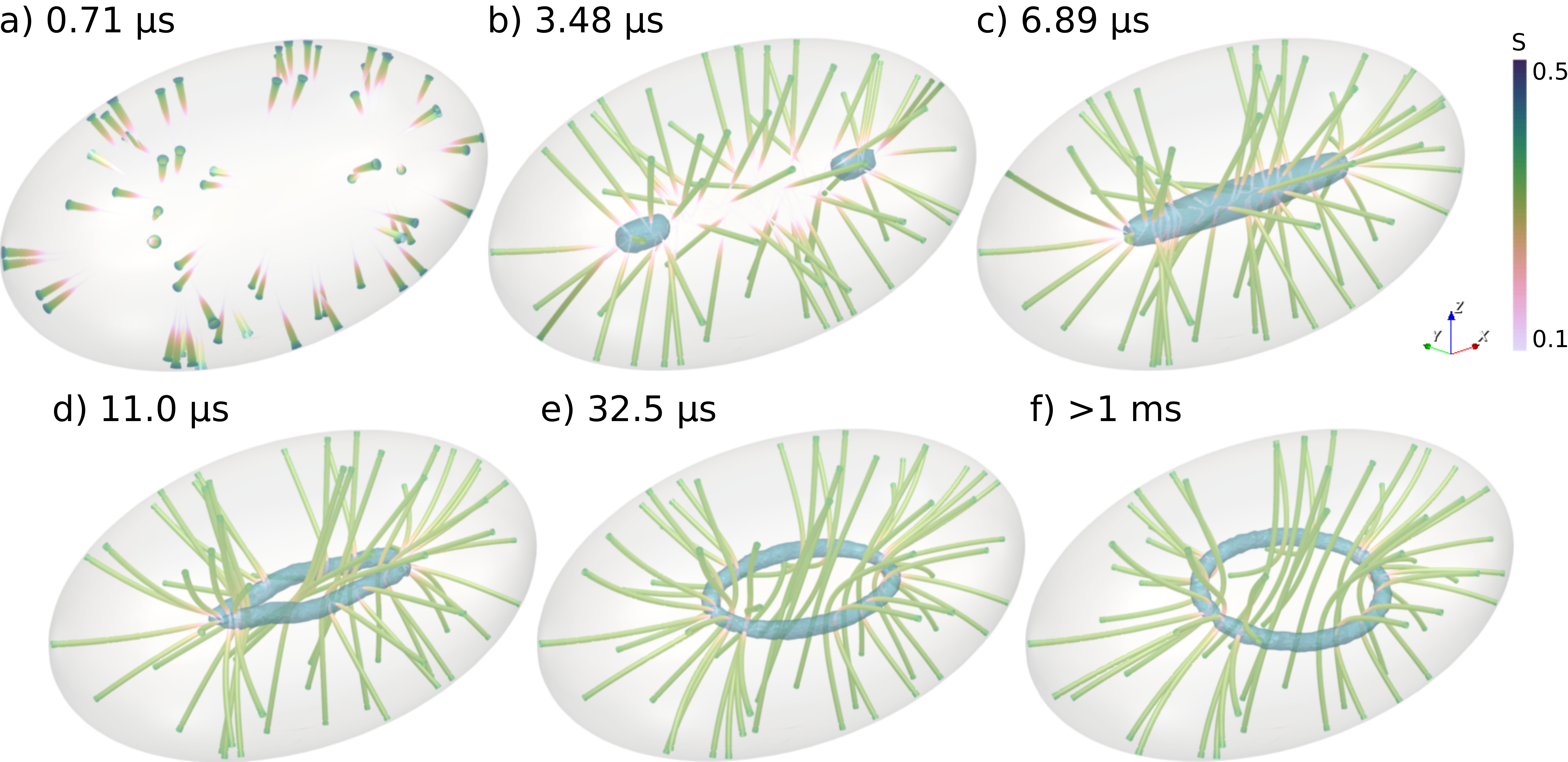}
    \caption{Simulation visualizations of the formation process of a prolate ($R = 2$) nematic droplet from an initially isotropic (disordered) state. Hyperstreamlines colored by the magnitude of the uniaxial nematic scalar order parameter $S$ are used to visualize nematic orientation (alignment tensor) and isosurfaces indicate nanoscale defect ``core'' regions.}
    \label{fig:formation_prolate}
\end{figure}

However, prolate nematic droplets behave differently.
While prolate spheroids can also be generated by extruding an ellipse, the homeotropic surface anchoring conditions distort the symmetric nature of the system.
Figure \ref{fig:formation_prolate} shows the formation dynamics of a prolate nematic droplet, which is found to exhibit the same general regimes as the oblate droplet: free growth (Figure~\ref{fig:formation_prolate}a), defect formation (Figures~\ref{fig:formation_prolate}b--d), and bulk relaxation (Figures~\ref{fig:formation_prolate}e,f).

Despite being topologically equivalent to the oblate droplet, the defect formation mechanism for a prolate droplet is substantially more complex.
First, a pair of $+1$ point defect-like structures form as the high-curvature focal regions impinge (Figure \ref{fig:formation_prolate}b).
These structures are not true point defects in that the nematic phase within the droplet is not fully-formed.
The defect formation mechanism proceeds through the continued impingement of the isotropic/nematic interface along the droplet equator.
This results in the point-like defects growing into the center of the droplet where they impinge to form a high-energy $+1$ disclination line.
As expected based on past two-dimensional simulation results \cite{Sonnet2001,Yan2002} and defect energy scaling analysis \cite{Kleman1982}, this defect line then splits into a $+\frac{1}{2}$ disclination loop for the prolate and spherical droplet (not shown) cases.
Notably, the simulations predict the dynamic mechanism through which this transition occurs.
Figures \ref{fig:formation_prolate}c,d show that there is a degeneracy in the direction in which the $+1$ disclination line splits, which results in this splitting direction varying along its length.
The resulting $+\frac{1}{2}$ disclination loop then undergoes an elastic relaxation process driven by defect line tension, bending, and torsion.

\begin{figure}
	\centering
	\includegraphics[width=0.65\linewidth]{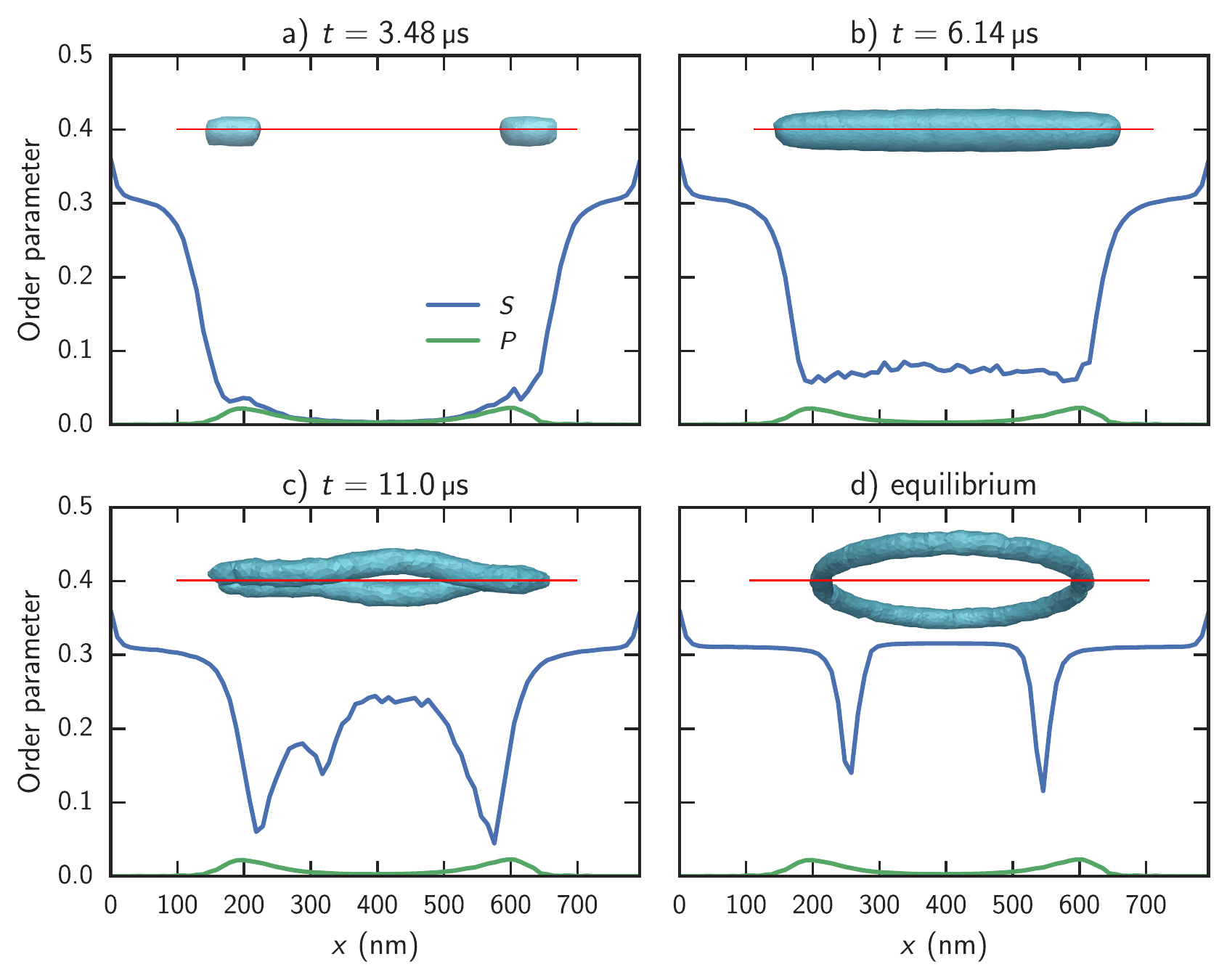}
	\caption{Plot of uniaxial $S$ and biaxial $P$ nematic order parameters versus position along the major axis of the $R=2$ droplet (illustrated in red) showing the progression of $+1$ disclination formation and splitting process.}
	\label{fig:defect_splitting}
\end{figure}

Figure \ref{fig:defect_splitting} shows the uniaxial $S$ and biaxial $P$ nematic order parameters in the vicinity of the central region of the prolate droplet during the disclination splitting process.
This process is similar to disclination line-loop dynamics observed by Shams and Rey \cite{Shams2012,Shams2015}, for which they developed a nematic elastica model for defect dynamics which captures line tension and bending of disclinations.
In the presently observed defect splitting process, line torsion, in addition to tension and bending, would need to be accounted for which could be accomplished through incorporating higher order terms in the nematic elastica model.
Figures \ref{fig:defect_splitting}a--b show the formation of an unstable $+1$ disclination line originating from the joining of a pair of $+1$ disclination lines growing into the unstable isotropic center of the prolate droplet.

Figure \ref{fig:defect_splitting}b shows that the central region of the fully-formed $+1$ disclination line is uniaxial, in agreement with past theoretical predictions \cite{Kralj1999}.
Figure \ref{fig:defect_splitting}c shows the initial distorted $+\frac{1}{2}$ disclination loop immediately following the splitting process.
The loop has significant bending and torsion resulting from the degeneracy in the splitting process.
It eventually relaxes into a loop with no torsion (Figure \ref{fig:defect_splitting}d) where the central region of the droplet is well-aligned with little distortion of the nematic director.

Finally, following the complex defect formation regime, the bulk relaxation regime is observed where the fully-formed nematic texture of the droplet relaxes through bulk reorientation and expansion of the disclination loop.
Comparing the equilibrium nematic textures of the oblate (Figure \ref{fig:formation_oblate}f) and prolate droplets (Figure \ref{fig:formation_prolate}f), the oblate droplet exhibits a relatively uniform nematic texture due to the surface exhibiting commensurate anchoring conditions with the bulk elasticity.
The prolate droplet exhibits a more non-uniform texture, and is more similar to the radial-like textures often observed in spherical nematic droplets.

\section{EXTERNAL FIELD-DRIVEN REORIENTATION AND RELAXATION}\label{sec:res:switching}

Electric-field driven reorientation of nematic droplets is a key process in the operation of PDLC-based technology.
Past experimental research has shown that droplet shape has a significant effect on the electro-optical switching process and can result in shorter switching times \cite{Klosowicz2005,Wu1989}.
Subsequently, simulations were performed for both oblate and prolate spheroids using the equilibrium states resulting from the formation process (Figures \ref{fig:formation_oblate}f and \ref{fig:formation_prolate}f, respectively).
Electric fields ranging from \SI{0}-\SI{14}{\volt\per\micro\meter} were applied in the direction parallel to the major axis of the droplets ($x$-axis), which, for the case of film stretching, corresponds to the direction orthogonal to the optical axis of the droplet at equilibrium.
Since 5CB is a positive dielectric anisotropy LC, nematic orientation parallel to the electric field is energetically favored.
Thus, imposition of the electric field orthogonal to the optical axis results in the maximum amount of field-induced reorientation, leading to more complex and interesting dynamics.
This corresponds to an in-plane switching mode that has been explored experimentally for PDLC-based devices \cite{Drevensek-Olenik2006,Park2015}.

Two different field-switching regimes were observed depending on the magnitude of the electric field, corresponding to a Fredericks-like transition.
For electric fields strengths $E$ below a critical value $E_c$, the nematic texture changes only slightly without undergoing bulk reorientation in the field direction.
In contrast, for $E$ above $E_{c}$, a complex reorientation process occurs with defect dynamics and a reorientation of the nematic texture to a field-aligned state.
Overall, a general sequence of three dynamic regimes, consistent with nematic capillaries \cite{Khayyatzadeh2015}, can be identified during this process, consisting of:
\begin{description}
	\item[Regime I.] \emph{Bulk growth and recession}, involving growth of the field-aligned focal regions and recession of the misaligned central region;
	\item[Regime II-A.] \emph{Disclination and bulk rotation}, involving rotation of the disclination loop orthogonal to the field direction; and
	\item[Regime II-B.] \emph{Bulk relaxation}, involving expansion of the disclination loop until the force from the applied field equilibrates with the elastic and anchoring forces in the system.
\end{description}
In all cases it was observed that upon release of the electric field, the nematic texture was restored to the initial equilibrium texture resulting from the earlier formation process.

\begin{figure}
	\centering
	\includegraphics[width=0.98\linewidth]{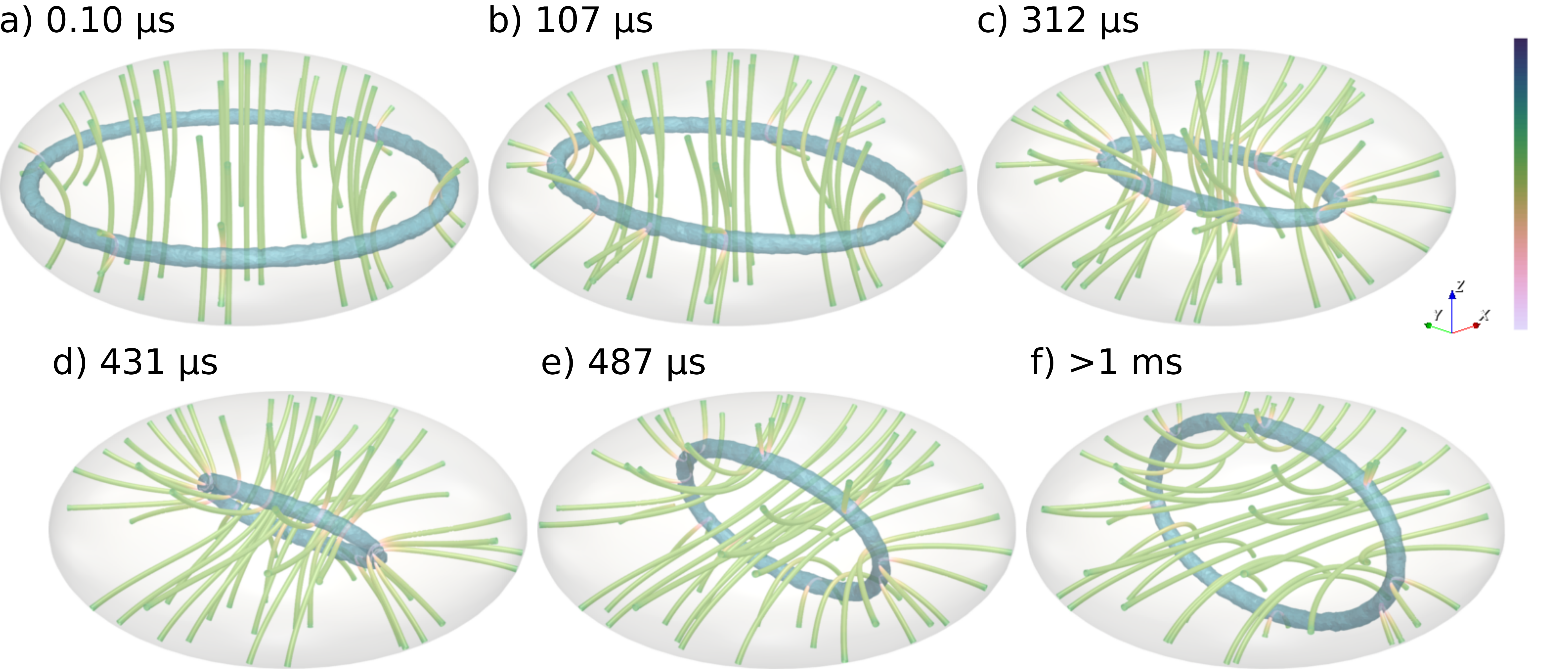}
	\caption{(a-f) Simulation visualizations of the electric field-switching process for $E = \SI{14}{\volt\per\micro\meter} > E_c$ applied along the $x$-axis of an oblate ($R = 0.5$) nematic droplet starting from (a) the equilibrium texture (following formation) and proceeding to the (f) the field-driven equilibrium texture.  Hyperstreamlines colored by the magnitude of the uniaxial nematic scalar order parameter $S$ are used to visualize nematic orientation (alignment tensor) and isosurfaces indicate nanoscale defect ``core'' regions.}
	\label{fig:switching-oblate}
\end{figure}

Figure~\ref{fig:switching-oblate} shows simulation results of the field-driven switching dynamics of a $R=0.5$ oblate nematic droplet where $E > E_c$.
Initially, the disclination loop contracts along the $x$-axis (Figures~\ref{fig:switching-oblate}b--c, dynamic regime I) as the field-aligned regions grow.
This process ends once the defect loop is ``compressed'' sufficiently into an elliptic shape such that the elastic energy penalty resulting further shape dynamics of the defect loop approaches that of the applied field.
For oblate droplet simulations where $E < E_c$ (not shown), dynamic regime I was the only dynamic regime observed.

Figures~\ref{fig:switching-oblate}d--e show the disclination/bulk rotation regime that follows.
Unlike the dynamic mechanism for field-switching of nematic capillaries \cite{Khayyatzadeh2015}, the rotation of the loop is accompanied by both expansion of the loop and bulk rotation of the nematic director throughout the droplet.
This corresponds to a combination of dynamic regimes II-A and II-B.

\begin{figure}
   	\centering
   	\includegraphics[width=0.98\linewidth]{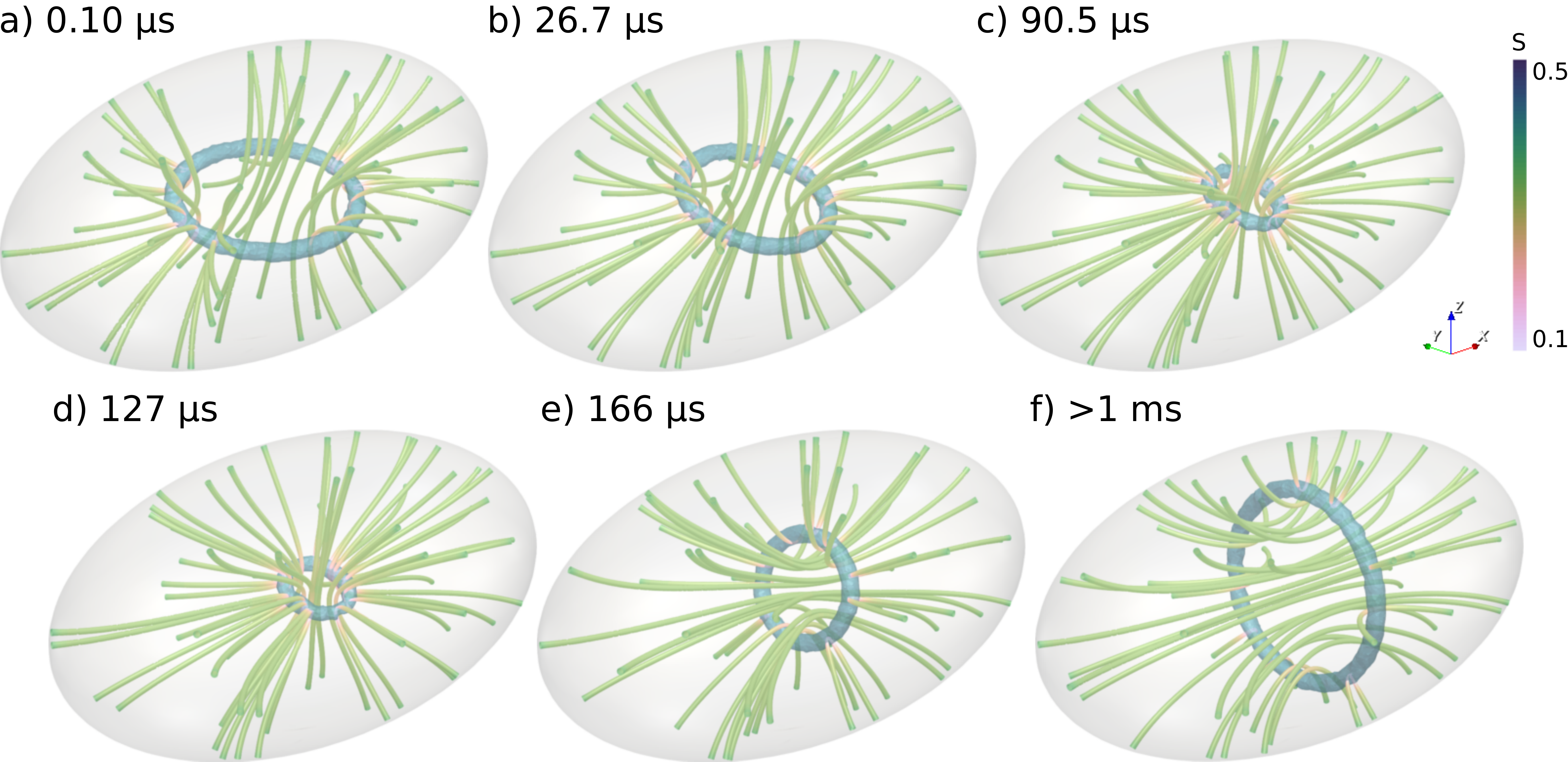}
    \caption{(a-f) Simulation visualizations of the electric field-switching process for $E = \SI{14}{\volt\per\micro\meter} > E_c$ applied along the $x$-axis of a prolate ($R = 2$) nematic droplet starting from (a) the equilibrium texture (following formation) and proceeding to the (f) the field-driven equilibrium texture.  Hyperstreamlines colored by the magnitude of the uniaxial nematic scalar order parameter $S$ are used to visualize nematic orientation (alignment tensor) and isosurfaces indicate nanoscale defect ``core'' regions.}
    \label{fig:switching-prolate}
\end{figure}

Figure~\ref{fig:switching-prolate} shows the field-driven switching dynamics of a $R=2$ prolate droplet where $E > E_c$.
The dynamic regimes observed here are more similar to those for nematic capillaries than for oblate spheroids.
In particular, the transition between dynamic regime II-A and II-B is more distinct.
This result can be attributed to the difference in disclination loop structure between the two cases, which is imposed by their geometries.
For the prolate droplet, as the size of the disclination loop decreases following application of the electric field, the loop becomes circular and its size is nanoscale.
In contrast, the oblate droplet disclination loop transitions from circular to elliptic after application of the field and the major axis of the elliptic loop maintains the micron-scale size of the overall droplet.
Next, dynamic regime II-A proceeds (Figures \ref{fig:switching-prolate}c--d) with a minimal increase in the defect loop diameter, unlike in the oblate case.
Following this, dynamic regime II-B is observed which the disclination loop diameter transitions from nanoscale to micron-scale, corresponding to the length scale imposed by the droplet geometry.

\begin{figure}
	\centering
	\includegraphics[width=0.98\linewidth]{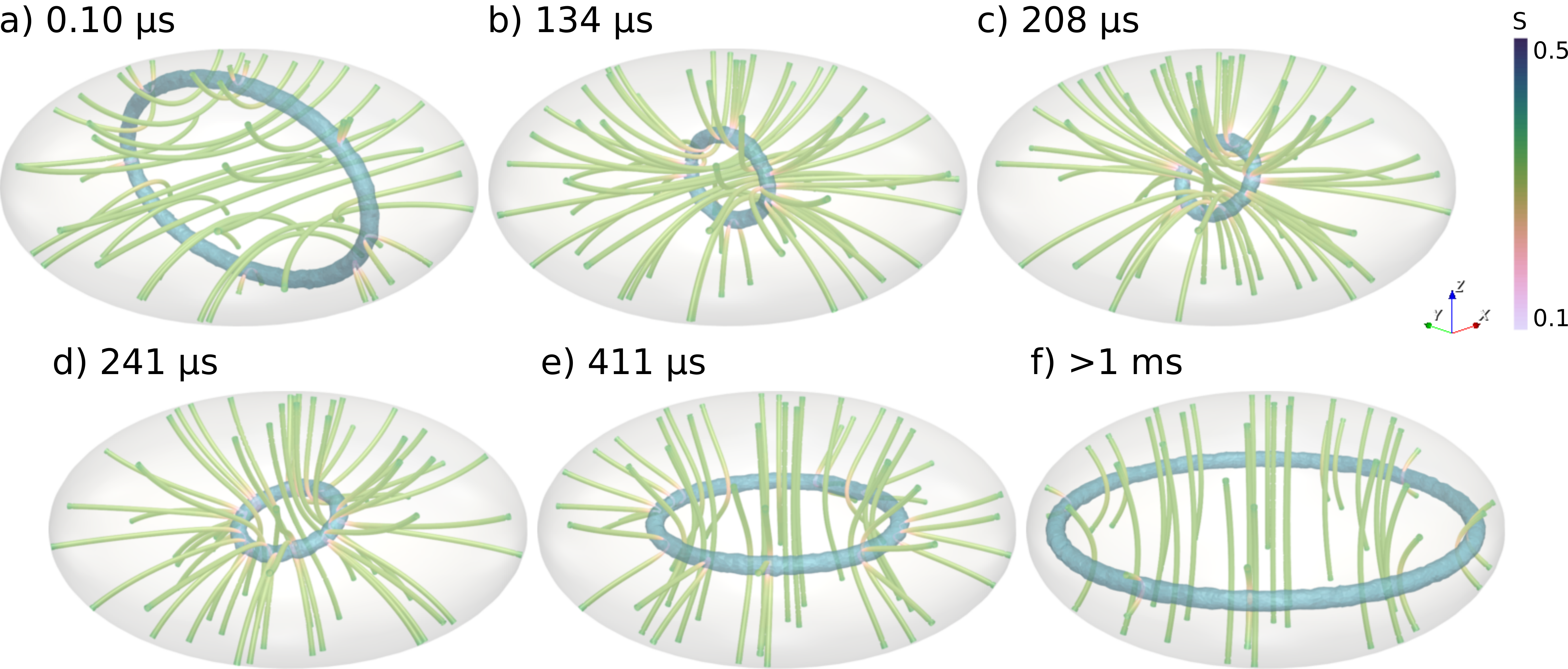}
	\caption{(a-f) Simulation visualizations of the field-off relaxation process after applying a field $E = \SI{14}{\volt\per\micro\meter} > E_c$ along the $x$-axis of an oblate ($R = 0.5$) nematic droplet starting from (a) the field-on equilibrium texture and proceeding to the (f) the field-off equilibrium texture.  Hyperstreamlines colored by the magnitude of the uniaxial nematic scalar order parameter $S$ are used to visualize nematic orientation (alignment tensor) and isosurfaces indicate nanoscale defect ``core'' regions.}
	\label{fig:relaxation-oblate}
\end{figure}

\begin{figure}
	\centering
	\includegraphics[width=0.98\linewidth]{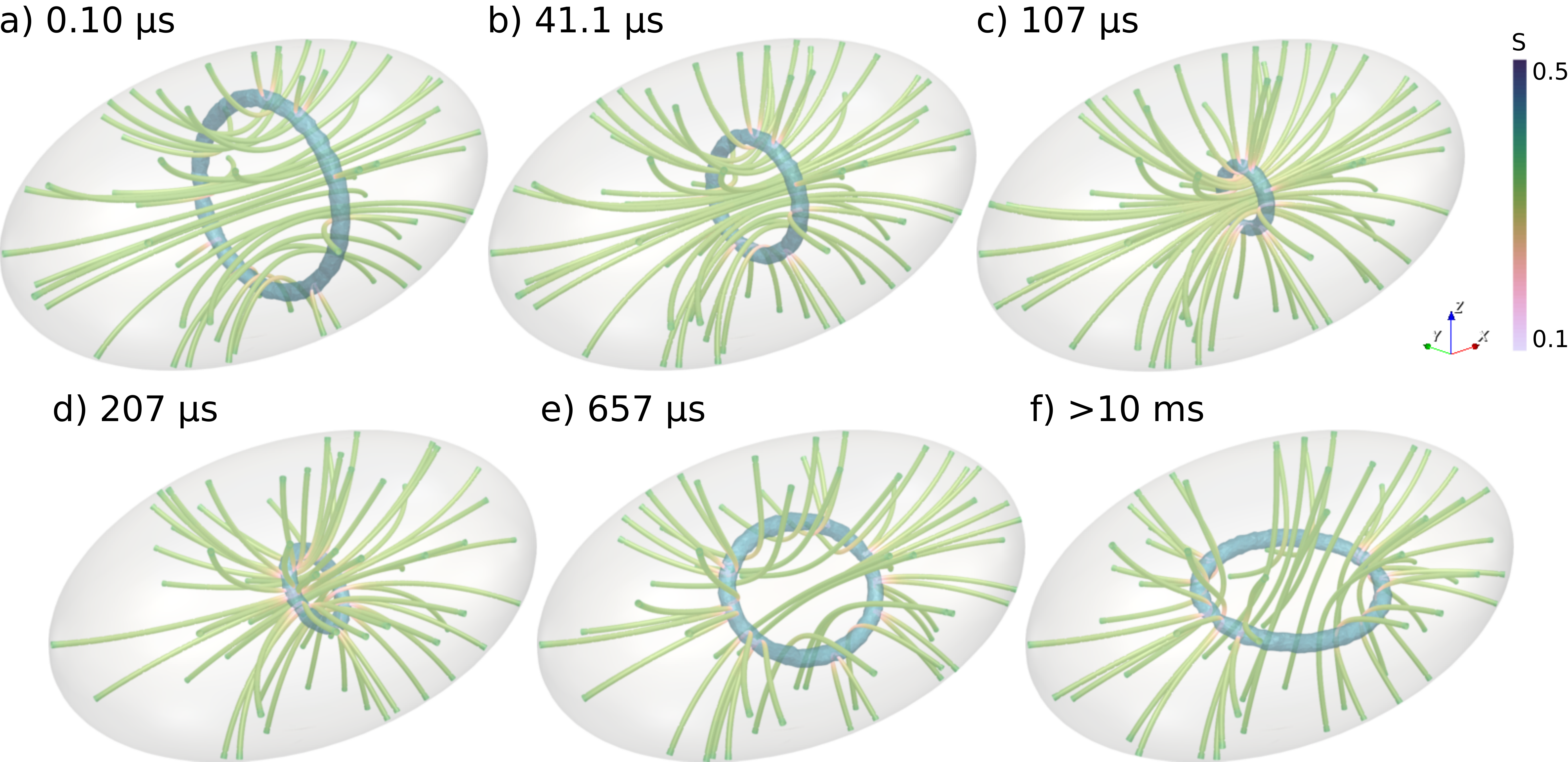}
	\caption{(a-f) Simulation visualizations of the field-off relaxation process after applying a field $E = \SI{14}{\volt\per\micro\meter} > E_c$ along the $x$-axis of a prolate ($R = 2$) nematic droplet starting from (a) the field-on equilibrium texture and proceeding to the (f) the field-off equilibrium texture.  Hyperstreamlines colored by the magnitude of the uniaxial nematic scalar order parameter $S$ are used to visualize nematic orientation (alignment tensor) and isosurfaces indicate nanoscale defect ``core'' regions.}
	\label{fig:relaxation-prolate}
\end{figure}

Upon release of the external field, the nematic texture at equilibrium while the field was applied is now a high-energy state.
Relaxation of the texture back to the original equilibrium state is due to a so-called ``restoring'' force \cite{Drzaic1988} which arises from a combination of confinement geometry and surface anchoring conditions.
Figures~\ref{fig:relaxation-oblate}--\ref{fig:relaxation-prolate} show simulation results of these dynamic mechanisms for $R = 0.5$ oblate and $R = 2$ prolate droplets, respectively.
Here, the relaxation process is, qualitatively, the reverse of the field-on process.
One significant difference was observed for the oblate droplet, in which the shape of the disclination loop during the relaxation process is different than that for the field-on case.

Focusing on the oblate droplet, the disclination loop shape during field-on conditions (Figure \ref{fig:switching-oblate}c) is elliptic while during release conditions (Figure \ref{fig:relaxation-oblate}b) it is circular.
For the field-on case, the elliptic disclination loop has a minor axis parallel to the field direction.
This elliptic shape is initially driven by the growth of the field-aligned regions and recession of the unaligned central region within the droplet.
As dynamic regime II-A proceeds, the elliptic character of the disclination loop is enhanced due to its proximity to the droplet's elliptic cross-section.
For the release case, the disclination loop is circular at the beginning of the rotation regime, and continues to maintain this shape throughout the rotation process.
As the disclination loop recedes from the elliptic part of the nematic/solid interface, it continuously transitions toward a state of minimum mean curvature which results in a circular shape.
As the loop rotates, this circular character of disclination loop is enhanced due to its proximity to a circular cross-section of the nematic/solid interface.
Additionally, unlike in the field-driven case there is a distinct transition from dynamic regime II-A (Figures \ref{fig:relaxation-oblate}b-d) to the dynamic regime II-B (Figures~\ref{fig:relaxation-oblate}e--f).

Figure~\ref{fig:relaxation-prolate} shows simulation results of the relaxation of a $R = 2$ prolate droplet.
For this case, the difference in the disclination loop shape between the field-on (Figure~\ref{fig:switching-prolate}d) and release (Figure~\ref{fig:relaxation-prolate}c) is more subtle, but similar to the oblate case.
For the field-on case, shown in Figure~\ref{fig:switching-prolate}d, the disclination loop is slightly elliptic with minor axis parallel to the field direction.
As it rotates the disclination loop transitions to a circular shape resulting from its proximity to a circular cross-section of the nematic/solid interface, shown in Figure~\ref{fig:switching-prolate}f.
Upon relaxation of the field (Figure~\ref{fig:relaxation-prolate}c), the disclination loop adopts a circular shape throughout the rotation process which is followed by transition to an elliptic shape due to its proximity to the elliptic cross-section of the droplet.

\section{DROPLET-SCALE DYNAMICS}

In order to analyze the external field-switching and relaxation dynamics quantitatively, a volume-averaged \emph{droplet} uniaxial scalar order parameter $S_d$ and director $\bm{n}_d$ can be determined \cite{Kelly1994} through eigendecomposition of the volume-averaged alignment tensor $\bm{Q}_{d}$:
\begin{equation}
    {Q_{d,ij}} = V^{-1}\int_{V} {Q_{ij}} \, dV
\end{equation}
where $V$ is the volume of the domain.
The droplet scalar order parameter $S_d$ is analogous to nematic scalar order parameter $S$ in Equation~\ref{eqn:qtensor}, where $S_d \rightarrow 0$ corresponds to a nematic droplet with no preferred alignment and $S_d\rightarrow 1$ corresponds to uniform aligned along $\bm{n}_d$.
The case where $S_d \rightarrow 0$ may correspond to two possible states of the nematic droplet: a fully isotropic (disordered) state or a  symmetrically radial nematic texture.
In this work, all analysis is performed for fully-formed nematic droplets and thus $S_d \rightarrow 0$ corresponds to the latter state.
From a general optical applications perspective, lower values of $S_d$ correspond to nematic droplets which scatter light, while higher values of $S_d$ correspond to nematic droplets with improved optical transparency \cite{Bloisi1996}.

\begin{figure}
	\centering
	\begin{subfigure}[b]{0.55\linewidth}
	\centering
	\includegraphics[width=\linewidth]{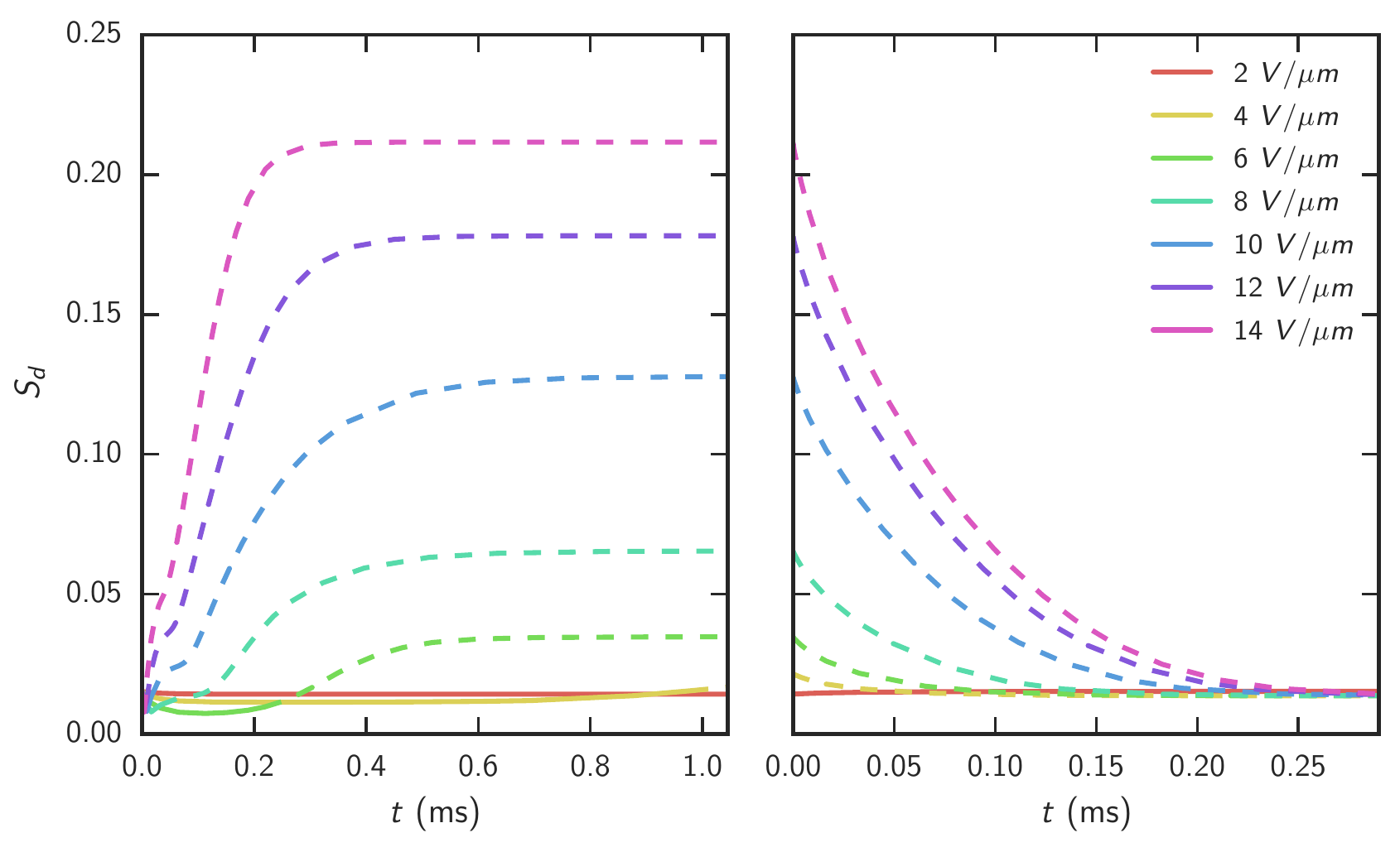}
	\caption{}\label{fig:sphere_Sd}
	\end{subfigure}
	\begin{subfigure}[b]{0.55\linewidth}
	\centering
	\includegraphics[width=\linewidth]{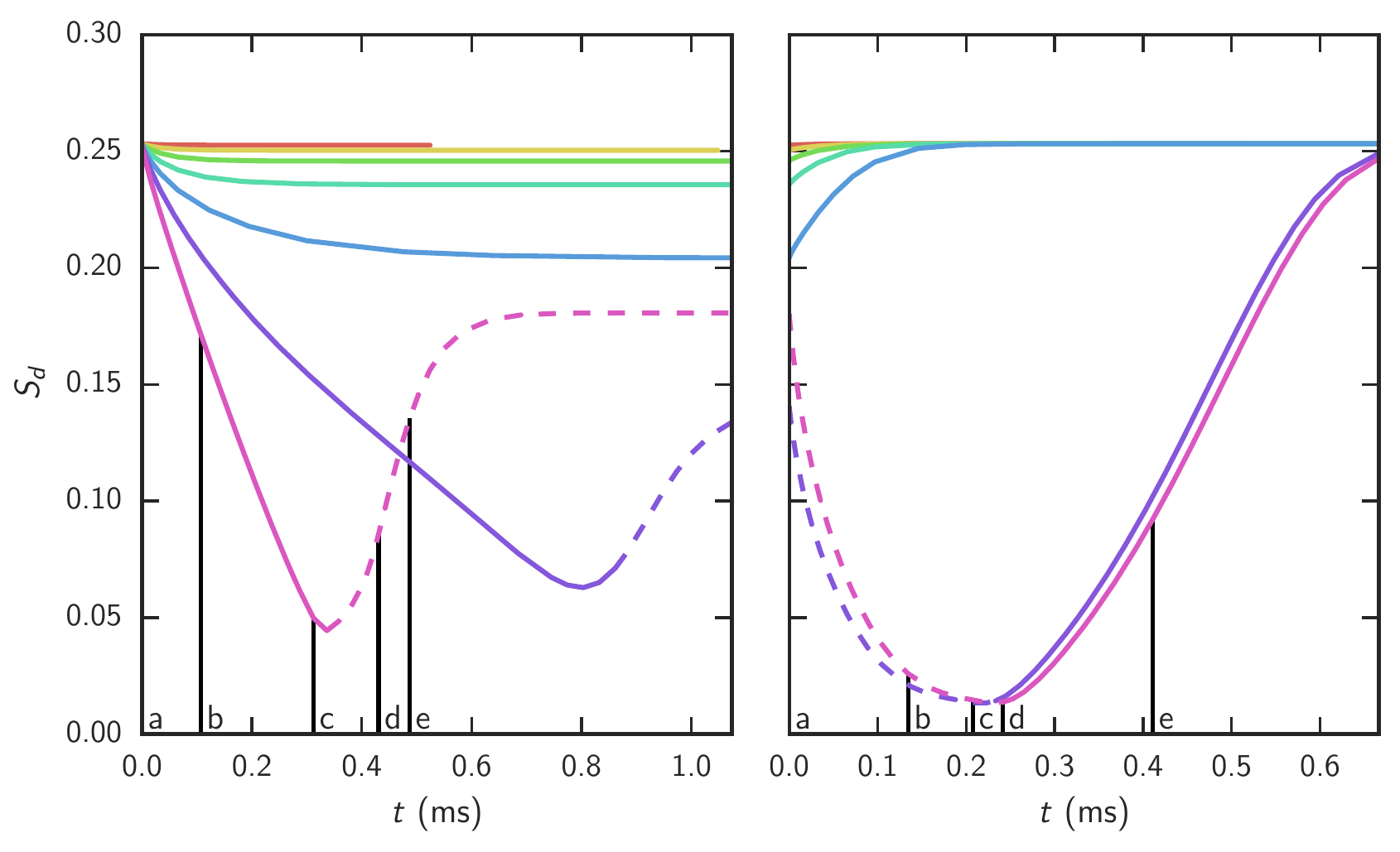}
	\caption{}\label{fig:oblate_Sd}
	\end{subfigure}
	\begin{subfigure}[b]{0.55\linewidth}
	\centering
	\includegraphics[width=\linewidth]{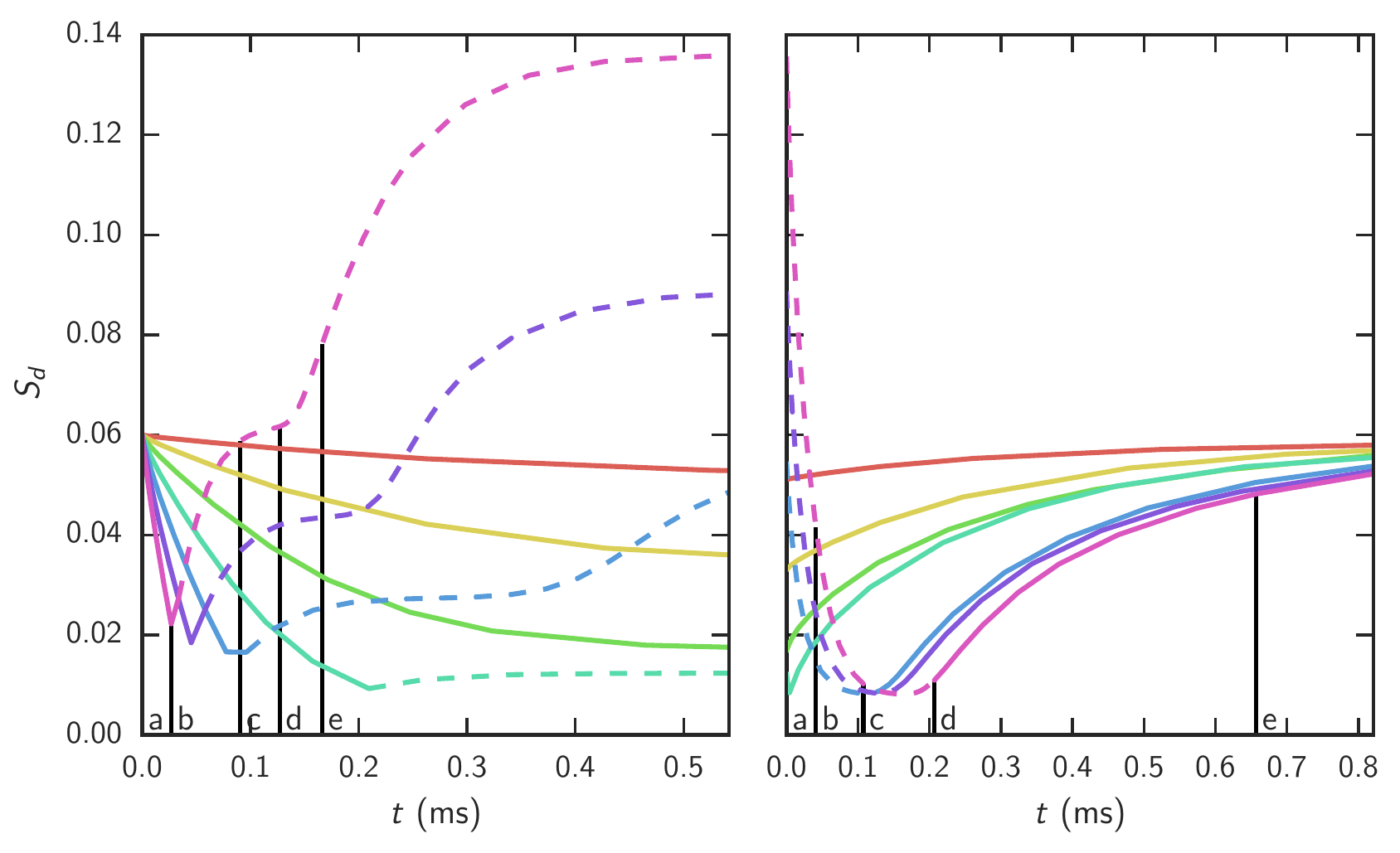}
	\caption{}\label{fig:prolate_Sd}
	\end{subfigure}
	\caption{Droplet-scale order evolution plots for (a) $R\approx 1$ spherical (not shown), (b) $R = 0.5$ oblate (Figures \ref{fig:switching-oblate} and \ref{fig:relaxation-oblate}), and (c) $R=2$ prolate (Figures \ref{fig:switching-prolate} and \ref{fig:relaxation-prolate}) nematic droplets resulting from application (left column) and release (right column) of electric fields with strengths ranging from $E = 2-14 \si{\volt\per\micro\meter}$. Curves represent the droplet scalar order parameter $S_d$ with solid/dotted lines corresponding to the droplet director $\bm{n}_d$ orthogonal/parallel to the electric field direction. Vertical bars with labels indicate the simulation time at which the corresponding simulation snapshots were taken for the oblate (Figures \ref{fig:switching-oblate} and \ref{fig:relaxation-oblate}) and prolate (Figures \ref{fig:switching-prolate} and \ref{fig:relaxation-prolate}) switching dynamics.}
	\label{fig:droplet-switching-plot}
\end{figure}

Figure \ref{fig:droplet-switching-plot} shows the evolution of $S_d$ and $\bm{n}_d$ for the field-switching and relaxation simulations of oblate (Figures~\ref{fig:switching-oblate} and \ref{fig:relaxation-oblate}), prolate (Figures~\ref{fig:switching-prolate} and \ref{fig:relaxation-prolate}), and (not shown) spherical nematic droplets which were presented in the previous section.
For the field-on dynamics, evolution of $S_d$ for $E < E_c$ exhibits a single bulk growth/recession regime.
For the $E > E_c$ cases, however, the evolution of $S_d$ for spherical and prolate droplets is found to involve three dynamic regimes, while the oblate droplet involves only two.
These quantitative findings support the qualitative observations from the previous section, where for oblate droplets dynamic regimes II-A and II-B occur simultaneously, whereas for the prolate droplets they are distinct.
Furthermore, the spherical droplet field-switching dynamics are found to be comparable to that of the prolate droplet, except for that the field-alignment of the droplet director $\bm{n}_d$ occurs very early in the field-on process for the spherical case.

These trends also indicate that dynamic regime I for spherical and prolate droplets occurs in two stages, unlike for the oblate droplet case.
In Figures~\ref{fig:sphere_Sd} and \ref{fig:prolate_Sd} (spherical and prolate droplets), $S_d$ initially decreases during regime I, followed by a rotation of the droplet director $\bm{n}_d$ and a simultaneous increase in $S_d$.
This is more pronounced for the prolate droplet than the spherical droplet.
In Figure~\ref{fig:oblate_Sd}, $S_d$ does not exhibit this nonmonotonic evolution during dynamic regime I.


The difference in field-on dynamics between prolate/spherical and oblate droplets may be explained through qualitative comparison of the disclination dynamics of the prolate and oblate droplets during the initial bulk growth/recession regime.
Focusing on the prolate droplet case, during the first stage of this dynamic regime, the field-aligned regions of the droplet grow while simultaneously the disclination loop diameter decreases.
The decrease is disclination loop diameter does not initially result in interaction of adjacent regions of the loop, which would result in a high-energy elastic interaction of the nanoscale defect ``core'' regions \cite{Kleman1982}.
During the second stage of this regime, the droplet scalar order parameter evolution decreases resulting from an overall slowing of the reorientation dynamics.
This is due to a slowing down of the macroscale field-alignment in the bulk domain as the disclination loop diameter approaches a critical value where adjacent defect core regions interact \cite{Khayyatzadeh2015}.
Following this, the domain transitions to dynamic regime II-A which occurs rapidly followed by a long timescale regime II-B.
For the oblate droplet case, the dynamic regime I is not observed to have two stages, implying different dynamics of the disclination loop during this regime.
Referring back to Figure \ref{fig:switching-oblate}, as the disclination loop reduces in size, it forms an elliptic shape which results in the focal segments of the loop having high curvature.
These high-energy regions preclude the possibility of adjacent disclination cores approach each other, and thus dynamic regime I for the oblate droplet does not involve interaction of adjacent defect core regions of the loop, unlike in the prolate case.

As mentioned in the previous section, the field-off/release dynamics, also shown in Figure \ref{fig:droplet-switching-plot}, are inherently different from the field-on dynamics due to the absence of an external field.
The restoring force resulting from the frustration of the field-on nematic texture with respect to the combination of the geometry, surface anchoring conditions, and nematic elasticity is substantially different for the spherical droplet case compared to both the oblate and prolate droplets in that there is only a very weakly imposed droplet director due to the geometry being essentially isometric.
Thus the release dynamics for this droplet involve only a bulk relaxation of the nematic texture.
As was described in the previous section, oblate and prolate droplets exhibit dynamics qualitatively similar to the field-on case, except in reverse.
Analysis of the droplet order parameter evolution for the field-off case shown in Figure \ref{fig:droplet-switching-plot} indicates that the dynamics are qualitatively similar, but both prolate and oblate droplets exhibit only two dynamic regimes with dynamic regimes II-A and II-B combined.

\begin{figure}
	\centering
	\includegraphics[width=0.9\linewidth]{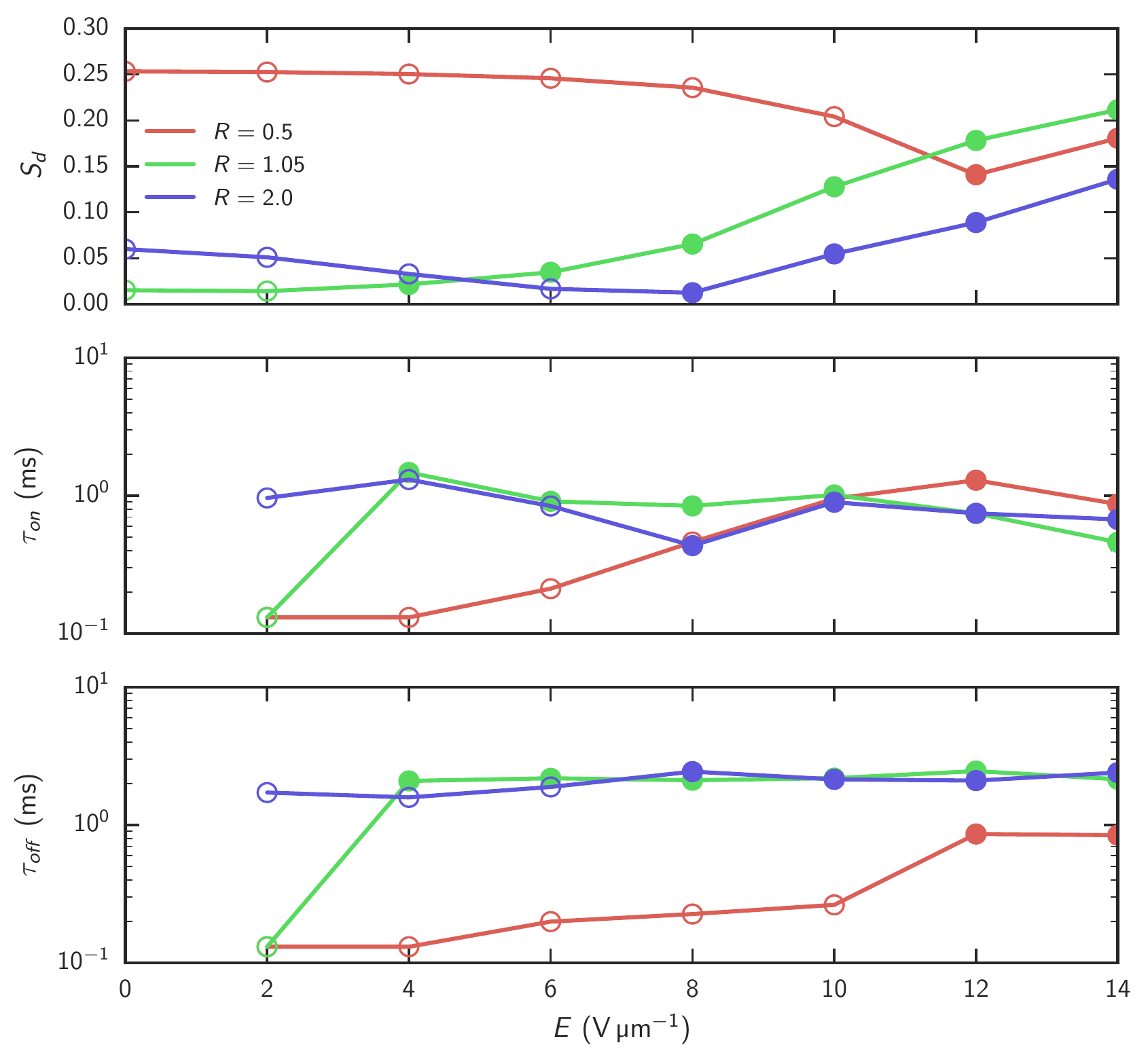}
	\caption{(top) Equilibrium droplet scalar order parameter $S_d$ versus electric field strength. (middle) Response times to reach field-driven equilibrium $\tau_{\mathrm{on}}$ versus electric field strength. (bottom) Response times to reach field-release equilibrium $\tau_{\mathrm{off}}$  versus electric field strength. Unfilled points correspond to droplet textures that are not field-aligned, while filled points correspond to those which are. }\label{fig:response_time}
\end{figure}

Equilibrium droplet scalar order parameter values and response times for a range of electric field strengths were also determined from simulations, which are of interest for PDLC-based devices and other technological applications.
Figure~\ref{fig:response_time} shows simulation results of droplet order parameter $S_d$ at equilibrium, field-on response times $\tau_{\mathrm{on}}$, and field-off response times $\tau_{\mathrm{off}}$ for oblate, spherical, and prolate droplets for a range of electric field strengths.
Measurements for $\tau_{\mathrm{on}}$ and $\tau_{\mathrm{off}}$ were estimated based on the time for $S_d$ to reach steady-state in order to be more comparable to experimental measurements, which are based on changes in optical film transmission \cite{Erdmann1989}.

As shown in Figure \ref{fig:response_time}, equilibrium $S_d$ values varied significantly  depending on both droplet shape and field strength.
Spherical droplets, which exhibit the lowest $E_c$, lack a strongly preferred droplet director, meaning that even relatively weak electric fields are effective for field-aligning the nematic texture.
Furthermore, the droplet order parameter $S_d$ increases monotonically with increasing field strength.
In contrast, for both oblate and prolate droplets, $S_d$ is nonmonotonic with respect to electric field strength, initially decreasing for $E < E_c$ and then increasing as $E > E_c$.
For the cases where $E < E_c$, oblate and prolate droplet responses do not involve reorientation of the droplet director.
Instead, $S_d$ decreases corresponding to decreased nematic alignment about the intrinsic droplet director resulting from geometry and anchoring conditions.
For the cases where $E > E_c$, full droplet director reorientation occurs in both prolate and oblate droplets, but to differing degrees.
The critical field strength for the oblate droplet reorientation is relatively high (\numrange[range-phrase = --]{10}{12} \si{\volt\per\micro\meter}), due to the large portion of the nematic/solid interface promoting alignment along the intrinsic droplet director.
In contrast, the critical field strength for the prolate droplet rorientation is relatively low (\numrange[range-phrase = --]{6}{8} \si{\volt\per\micro\meter}) for the opposite reason.
Once reorientation occurs, the droplet scalar order parameter increases linearly with $E$ as the electric field influence overcomes surface anchoring forces.


The results for field-on response times for both prolate and spherical droplets are comparable to experimental results for spherical droplets under similar conditions ($\approx \SI{1}{\milli\second}$) \cite{Xie2004}.
Both field-on and field-off response times for oblate droplets are significantly lower, on the order of $\SI{0.1}{\milli\second}$, which is due to their negligible change in texture in response to an applied field, as indicated by very little change in the droplet order parameter between field-on and field-off states.
However, simulation results for field-off response times for both prolate and spherical droplets are somewhat lower than experimental results \cite{Xie2004}, $\approx \SI{10}{\milli\second}$ versus $\approx \SI{30}{\milli\second}$, respectively.
This can be attributed to the significantly larger length scale of nematic droplets studied experimentally, $1-7\si{\micro\meter}$, which results in a decreased ratio of restoring to viscous forces, slowing down droplet dynamics.

\section{CONCLUSIONS}

In this work continuum simulations were performed in order to predict the dynamic mechanisms involved in the formation, field switching, and relaxation of nematic LC droplets with varying spheroidal geometry.
The presented simulation results have both fundamental and technological relevance in that formation and field-switching dynamic mechanisms were previously poorly understood and of significant relevance to the performance of PDLC-based optical functional materials.
The key feature of these nematic domains is the presence of nanoscale defect structures which contribute to the dynamics of the micron-scale domain in complex ways.

Simulations of formation dynamics from an initially unstable isotropic phase predict intrinsically different defect formation mechanisms in anisometric droplets (oblate and prolate) compared to spherical ones.
Defect loop structures, which are topologically imposed by domain geometry and anchoring conditions, are observed to form through the combination of defect shedding and splitting dynamic mechanisms.
A degeneracy in the splitting of a $+1$ disclination line structure into a $+\frac{1}{2}$ disclination loop is predicted to result in an ``unraveling'' of the nanoscale loop structure, similar to the nematic elastica behavior observed in nematic capillaries.

Simulations of electric field-driven reorientation and relaxation dynamics reveal the mechanisms of the reorientation process, which are highly dependent on domain shape and external field strength.
Both oblate and prolate spheroidal droplets are found to have qualitatively similar dynamic reorientation mechanisms, with the critical (reorientation) electric field strength $E_c$ being significantly higher than for spherical droplets.
For electric fields $E < E_c$, the nematic texture of anisometric droplets becomes increasingly frustrated between the orientation imposed by the external field and that preferred by the geometry and anchoring conditions.
This corresponds to an optical state that is increasingly light scattering.
For electric fields $E > E_c$, the nematic texture transitions to a field-aligned state through a series of complex and distinct dynamic mechanisms involving both micron-scale reorientation and nanoscale defect dynamics.

In summary, the presented results provide both qualitative and quantitative insight into the dynamics of nematic spheroids with resolution of the nanoscale length and timescales inherent to LC domains which include defects.
These simulations include the dynamic regimes relevant to PDLC-based devices and thus could be used to guide the design and optimization of their performance as optical functional materials.
Additionally, these results provide fundamental insight into the effects of nanoscale defect dynamics on confined LC domains.

\section{METHODS}\label{sec:methods}

\paragraph{Nematic Reorientation Dynamics Model.}

Simulations are performed using the Landau--de Gennes continuum model for the nematic phase \cite{deGennes1995, Barbero2001}, which uses an alignment tensor \cite{Sonnet1995}, or $Q$-tensor, order parameter to quantify nematic order:
\begin{equation}\label{eqn:qtensor}
    Q_{ij} = S(n_i n_j - \frac{1}{3}\delta_{ij}) + P(m_i m_j - l_i l_j)
\end{equation}
where $S$ and $P$ are uniaxial and biaxial nematic scalar order parameters, $n_i$ is the nematic director, and $m_i$, $l_i$ are the biaxial orientation vectors.
The Helmholtz free energy density of the domain is \cite{deGennes1995, Barbero2001}:
\begin{multline}
f_{b} -f_{iso} = \frac{1}{2}a(Q_{ij}Q_{ji}) + \frac{1}{3}b(Q_{ij}Q_{jk}Q_{ki}) + \frac{1}{4}c(Q_{ij}Q_{ji})^2 \\
+ \frac{1}{2}L_1(\partial_i Q_{jk} \partial_i Q_{kj}) + \frac{1}{2}L_2(\partial_i Q_{ij} \partial_k Q_{kj}) + \frac{1}{2}L_3(Q_{ij}\partial_iQ_{kl} \partial_jQ_{kl}) + \frac{1}{2}L_{24}(\partial_kQ_{ij} \partial_jQ_{ik}) \\
- \frac{\epsilon_{\circ}}{8\pi}\left[ \left( \frac{\epsilon_{\parallel}+2\epsilon_{\perp} }{3} \delta_{ij} +( \epsilon_{\parallel} -\epsilon_{\perp} ) Q_{ij} \right) \right] E_{j} E_{i}
\end{multline}
where $f_{iso}$ is the free energy of the isotropic phase, which is assumed to be constant.
All three second-order terms in $Q_{ij}$ are used, while the third-order $L_3$ term is used in order to resolve splay-bend anisotropy, and $L_{24}$ is used to quantify saddle-splay elasticity.
The $L_{24}$ term is also referred to as $L_3$ or $L_4$, depending on the reference source \cite{Barbero2001,Sonnet2012,Tomar2012}.

Additionally, a contribution to the free energy from the solid/nematic interface corresponding to homeotropic surface anchoring is used \cite{Barbero2005}:
\begin{equation}
f_{s} = \alpha k_{i} Q_{ij} k_{j}
\end{equation}
where $k_i$ is the surface unit normal and $\alpha$ is the surface anchoring strength.
A value of $\alpha = \SI{-1.0e-4}{J/m^2}$ was used, which is corresponds moderately strong surface anchoring with a surface extrapolation length $\xi_s = \frac{L_1}{\alpha} \approx \SI{100}{nm})$ \cite{Barbero2005,Ravnik2009}.
The total free energy of the domain includes both bulk and surface contributions:
\begin{equation}\label{eqn:total_free_energy}
F[Q_{ij}] = \int_{V} f_{b} \, dV + \int_{S} f_{s} \, d S.
\end{equation}

Nematic reorientation dynamics are modelled using the time-dependent Ginzburg-Landau model\cite{Hohenberg1977}:
\begin{equation}
    \frac{\partial Q_{ij}}{\partial t} = - \Gamma \left[\frac{\delta F}{\delta Q_{ij}}\right]^{ST}
\end{equation}
where $\Gamma = \mu_{r}^{-1}$ where $\mu_{r}$ is the rotational viscosity of the nematic phase, and $[ ]^{ST}$ indicates the symmetric-traceless component.

Numerical solution of the resulting system of nonlinear partial differential equations was performed using the finite element method with the software package FEniCS \cite{FEniCSBook} on meshes of spheroid geometries or ``droplets'' with aspect ratio $R = \frac{c}{a}$, where $c$ and $a$ correspond to the lengths of the major and minor axes of the spheroid.
Droplet volume was maintained constant for each geometry and set to be equivalent to the volume of a perfectly spherical droplet with diameter \SI{500}{\nm} with mesh spacing less than the nematic coherence length in order to accurately resolve the defect structure.

The governing equations were nondimensionalized before solving, which gives rise to a time scale $t_s$:
\begin{equation}\label{eqn:dimensionless}
    \tilde{t} = \frac{t}{t_s}, \quad t_s = \frac{\mu_r}{a_0 T_{ni}}
\end{equation}
An estimate for $t_s$ can be calculated using the parameters given in Table~\ref{tab:model_parameters}.

\paragraph{Simulation Method Conditions.}

The model parameters used approximate the liquid crystal 4-cyano-4'-pentylbiphenyl (5CB) and are given in Table~\ref{tab:model_parameters}.
Values were chosen according to experimental data for a temperature of $T = \SI{307}{K}$  \cite{Luckhurst2001,Bogi2001,Wincure2007a}.

\begin{table}
    \caption{Material parameters for 5CB.\label{tab:model_parameters}}
    \begin{tabular}{ccc}
    $T_{ni}$ & $307.35$ &  \si{K} \\
    $a_{0}$ & \num{1.4e6} & \si{J/m^3 K} \\
    $b$ & \num{1.8e6} & \si{J/m^3} \\
    $c$ & \num{3.6e6} & \si{J/m^3} \\
    $L_{1}$ & \num{9.1e-12} & \si{J/m} \\
    $L_{2}$ & \num{4.4e-12} & \si{J/m} \\
    $L_{3}$ & \num{7.1e-12} & \si{J/m} \\
    $L_{24}$ & \num{4.2e-12} & \si{J/m} \\
    $\epsilon_{\parallel}$ & $16.5$ &  (relative) \\
    $\epsilon_{\perp}$ & $8.2$ &  (relative) \\
	$\mu_r$ & $0.055$ & \si{Ns/m^2} \\
    \end{tabular}
\end{table}

The values of the elastic constants $L_1$ to $L_{24}$ were derived \cite{Mori1999,Tomar2012} from the Frank elastic constants $k_{11} = \SI{2.5e-12}{J/m}$, $k_{22} = \SI{1.7e-12}{J/m}$, and $k_{33} = \SI{3.0e-12}{J/m}$, which were determined using known empirical models \cite{Luckhurst2001,Bogi2001,Polak1994}.
The saddle-splay constant $k_{24}$, which has been difficult for researchers to measure consistently for 5CB \cite{Allender1991, Polak1994, Joshi2014}, was chosen such that the elastic energy penalty term $L_{24}$ remained positive ($k_{24} = 0.25k_{22}$).
However, this is not a strict condition and negative $L_{24}$ is possible as long as the Frank elastic constants satisfy Ericksen's inequalities \cite{Tomar2012}.

The simulation for determining field-off equilibrium droplet textures was initialized using a uniaxial boundary layer with scalar order parameter $S_0=S_{eq}$.
The boundary is aligned perpendicular to the surface in accordance with the surface boundary conditions ref. \citenum{Khayyatzadeh2015}.
Simulations of electric field switching were conducted using these equilibrium textures as initial conditions for each field strength studied.

\paragraph{Visualization.}

Three-dimensional visualizations of the droplets were generated using hyperstreamline seeding of the Q-tensor field \cite{Fu2015}.
Hyperstreamlines are used to represent the orientational order tensor $Q_{ij}(x,t)$ \cite{Delmarcelle1993,Fu2015}.
These structures are an extension of streamlines and orient along the director field $n_i(x,t)$, with varying width in order to visualize the additional degrees of freedom associated with tensorial data.
Hyperstreamlines are colored according to the scalar order parameter $S$.
Disinclination lines are indicated the blue contour surfaces which were computed for a fixed biaxial scalar order parameter $P > 0$.

\begin{acknowledgement}

This work was supported by the Natural Sciences and Engineering Research Council of Canada and Compute Canada.

\end{acknowledgement}

\begin{suppinfo}

\begin{itemize}
  \item \emph{sphere\_formation.mpg}: Video of formation dynamics for a $R \approx 1$ spherical droplet simulation. Hyperstreamlines colored by the magnitude of the uniaxial nematic scalar order parameter $S$ are used to visualize nematic orientation (alignment tensor) and isosurfaces indicate nanoscale defect ``core'' regions (refer to Methods section). Time is given as a dimensionless quantity (see eqn. \ref{eqn:dimensionless}).
  \item \emph{oblate\_formation.mpg}: Video of formation dynamics for a $R = 0.5$ oblate droplet simulation, corresponding to Figure~\ref{fig:formation_oblate}.
  \item \emph{prolate\_formation.mpg}: Video of formation dynamics for a $R = 2$ prolate droplet simulation, corresponding to Figure~\ref{fig:formation_prolate}.

  \item \emph{sphere\_fieldon\_14Vum.mpg}: Video of field-switching dynamics of a $R \approx 1$ spherical droplet for $E=\SI{14}{\volt\per\micro\meter} > E_c$ applied along the $x$-axis.
  \item \emph{oblate\_fieldon\_14Vum.mpg}: Video of field-switching dynamics of a $R = 0.5$ oblate droplet for $E=\SI{14}{\volt\per\micro\meter} > E_c$ applied along the $x$-axis, corresponding to Figure~\ref{fig:switching-oblate}.
  \item \emph{prolate\_fieldon\_14Vum.mpg}: Video of field-switching dynamics of a $R = 2$ prolate droplet for $E=\SI{14}{\volt\per\micro\meter} > E_c$ applied along the $x$-axis, corresponding to Figure~\ref{fig:switching-prolate}.

  \item \emph{sphere\_fieldrelease\_from\_14Vum.mpg}: Video of field-off relaxation dynamics for a $R \approx 1$ spherical droplet simulation.
  \item \emph{oblate\_fieldrelease\_from\_14Vum.mpg}: Video of field-off relaxation dynamics for a $R = 0.5$ oblate droplet simulation, corresponding to Figure~\ref{fig:relaxation-oblate}.
  \item \emph{prolate\_fieldrelease\_from\_14Vum.mpg}: Video of field-off relaxation dynamics for a $R = 2$ prolate droplet simulation, corresponding to Figure~\ref{fig:relaxation-prolate}.

\end{itemize}

\end{suppinfo}

\bibliography{self_assembly,computational,pdlcs}

\end{document}